\documentclass[aps,pre,epsf,twocolumn,floatfix]{revtex4-2}

\usepackage{amsmath}
\usepackage{amssymb}
\usepackage{epsfig}
\usepackage{graphicx}
\usepackage[T1]{fontenc}
\usepackage[utf8]{inputenc}
\usepackage{color}
\usepackage[normalem]{ulem}
\usepackage[colorlinks]{hyperref}
\usepackage{tikz}
\usetikzlibrary{positioning}

\usepackage{tikz}
\usetikzlibrary {shapes.geometric, calc}
\tikzset{
    my hex/.style={regular polygon, regular polygon sides=6, draw, inner sep=0pt, outer sep=0pt, minimum size=1cm},
    my circ/.style={draw, circle, fill=red!50!white, inner sep=0pt, minimum size=1.5mm}
}

\begin{document}

\title{Static versus dynamic universality in the site-diluted kagome Ising model}

\author{Alexandros Vasilopoulos}
\affiliation{School of Mathematics, Statistics and Actuarial Science, University of Essex, Colchester CO4 3SQ, United Kingdom}

\author{Zeynep Demir Vatansever}
\affiliation{Department of Physics, Dokuz Eyl\"{u}l University, TR-35160, Izmir, Turkey}

\author{Erol Vatansever}
\affiliation{Department of Physics, Dokuz Eyl\"{u}l University, TR-35160, Izmir, Turkey}

\author{Gerard T. Barkema}
\email{G.T.Barkema@uu.nl}
\affiliation{Department of Information and Computing Sciences, Utrecht University, Princetonplein 5, 3584 CC Utrecht, the Netherlands}

\author{Nikolaos G. Fytas}
\email{nikolaos.fytas@essex.ac.uk}

\affiliation{School of Mathematics, Statistics and Actuarial Science, University of Essex, Colchester CO4 3SQ, United Kingdom}
\affiliation{Section of Condensed Matter Physics, Department of Physics, National and Kapodistrian University of Athens, Panepistimiopolis, 157 72 Athens, Greece}

\date{\today}

\begin{abstract}
We study the site-diluted Ising model on the kagome 
lattice using the Wolff single-cluster algorithm, 
focusing on both equilibrium and dynamic critical 
properties. The equilibrium critical exponents 
$\nu = 1$ and $\gamma/\nu = 7/4$ retain their exact 
pure Ising values for all spin concentrations $p$ 
considered, consistent with the marginal irrelevance 
of disorder at $\alpha = 0$, with the specific heat 
crossing over from logarithmic to double-logarithmic 
growth upon dilution. These static results are shared 
between the kagome and square lattices. On the dynamic 
side, $z$ decreases upon dilution on both lattices---opposite to what is observed under local dynamics---with evidence of saturation to a $p$-independent 
value at strong dilution, suggesting a diluted dynamic 
fixed point. While $z$ is consistent between the two 
lattices in the pure case, it differs in the diluted 
regime, an effect we attribute to the two lattices 
sitting at different positions along the 
renormalization-group flow between the Ising and 
percolation fixed points, rather than to a true 
asymptotic breakdown of dynamic universality.
\end{abstract}

\maketitle

\section{Introduction}
\label{sec:intro}

Understanding the role of quenched disorder in shaping 
the critical behavior of magnetic systems is one of the 
fundamental questions of statistical mechanics. A 
particularly important aspect is whether the introduction 
of disorder modifies the universality class of a model, 
or whether the critical exponents remain robust against 
weak randomness. In this work we focus on the two-dimensional 
Ising model~\cite{mccoy:73, newman_book}, arguably the 
simplest and most studied model in statistical physics, 
for which the specific-heat exponent $\alpha = 0$ places 
the system at the marginal case of the Harris 
criterion~\cite{harris74}, where the criterion itself 
is inconclusive. More detailed renormalization-group 
calculations and extensive numerical computations 
establish that disorder is marginally irrelevant, 
leading not to a new universality class but to 
multiplicative logarithmic corrections to the pure 
fixed point, the so-called strong universality hypothesis~\cite{dotsenko83, shankar87, ludwig87, 
ludwig87b, ludwig90, wang90, shalaev94, ballesteros97, 
selke98, toldin08, kenna08, kenna09, fytas08, fytas10, 
dotsenko18}. While the static critical behavior of 
the disordered model is by now well understood, the 
dynamic critical behavior---and in particular its 
dependence on the type of Monte Carlo update, the 
nature of the disorder, and the underlying lattice 
geometry---remains significantly less 
explored, in both two and three dimensions~\cite{heuer93, hennecke93, janssen95, luo01, zheng02, zhong:20, kole22}.

The extension of universality concepts to dynamical 
processes~\cite{hohenberg:77} is less straightforward 
than in the static case, and lags behind its theoretical 
counterpart~\cite{hohenberg:77, folk:06}. Even for 
two-dimensional Ising model a rigorous analytical treatment of the critical dynamics remains elusive, and models within 
the same static universality class do not necessarily 
belong to the same dynamic universality 
class~\cite{bonati:25}. The dynamic critical exponent 
$z$, which governs the critical slowing down via the 
power-law relation $\tau \sim L^z$ at criticality, 
where $\tau$ denotes the autocorrelation time, depends 
sensitively on the choice of update scheme. For local 
(Metropolis) dynamics, $z \approx 2.16$ is well 
established~\cite{nightingale96, bisson25}, and this 
value is robust in two dimensions across different lattice geometries, consistent with the expectation that local dynamics 
are insensitive to the microscopic lattice structure 
at criticality. For cluster algorithms, the situation 
is fundamentally different: the Wolff~\cite{wolff89} 
and Swendsen-Wang~\cite{swendsen87} algorithms reduce 
$z$ dramatically relative to local updates in the 
pure case~\cite{baillie91,du06,liu14}, reflecting the collective nature of the cluster flip and its ability to 
decorrelate the system far more efficiently than 
single-spin updates.

The introduction of disorder modifies the dynamic 
critical behavior in a nontrivial and update-dependent 
way. The question of dynamic universality in disordered 
systems has a rich history. In a seminal contribution, 
Parisi \emph{et al.}~\cite{parisi99} studied the 
off-equilibrium critical dynamics of the 
three-dimensional diluted Ising model and demonstrated 
that the dynamic universality class is preserved under 
dilution for local dynamics, providing early numerical 
evidence for the robustness of dynamic universality 
in disordered systems. Subsequently, Hasenbusch and 
coauthors~\cite{hasenbusch07} established that the 
site-diluted, bond-diluted, and $\pm J$ versions of 
the three-dimensional Ising model share the same 
dynamic exponent for Metropolis updates, around 
$z = 2.35(2)$, consistent with dynamic 
renormalization-group predictions~\cite{janssen89}, 
while violations of dynamic universality have also 
been reported in other contexts~\cite{daSilva:09, 
zhong:20}. For cluster algorithms, the situation is 
qualitatively different: Ivaneyko \emph{et al.}~\cite{ivaneyko06} 
found that $z$ decreases for both Wolff and 
Swendsen-Wang updates upon site dilution of the 
three-dimensional Ising model, and recent studies 
have confirmed that bond disorder leads to a further 
decrease in $z$ under cluster dynamics in two 
dimensions~\cite{kanbur24, vatansever25}---in stark 
contrast to what is observed for local updates. 
Despite this progress, the dynamic critical behavior 
of the two-dimensional site-diluted Ising model 
under cluster dynamics has not been systematically addressed, 
and the dependence of $z$ on both the dilution 
strength and the underlying lattice geometry is not 
yet understood.

This work aims to fill this gap by addressing the 
following specific questions. Does site dilution 
reduce or increase $z$ under Wolff dynamics in two 
dimensions, and is this behavior consistent across 
different lattice geometries? Does $z$ depend on 
the dilution strength $p$, or does it saturate to 
a $p$-independent value for all $p < 1$, suggesting 
the existence of a diluted dynamic fixed point? And 
is the dynamic universality class shared between the 
kagome and square lattices, as is the case for the 
static universality class? Our results provide clear 
answers to all three questions: site dilution 
decreases $z$ under Wolff dynamics, consistent 
with previous results for bond disorder and in stark 
contrast to the behavior under local updates; the 
data are consistent with $z$ saturating to a 
$p$-independent value at strong dilution, suggesting 
a diluted dynamic fixed point; and while $z$ is 
shared between the kagome and square lattices in the 
pure case, it differs in the diluted regime, an effect 
we attribute to the two lattices sitting at different 
positions along the renormalization-group flow between 
the Ising and percolation fixed points, rather than 
to a true asymptotic breakdown of dynamic universality.

The remainder of the manuscript is organized as 
follows. In Section~\ref{sec:model} we define the 
model and describe the kagome lattice, motivating 
the choice of geometry in the context of static 
and dynamic universality. In Section~\ref{sec:numerics} 
we present an overview of our simulation approach, 
the protocols followed, and the observables of 
interest. Section~\ref{sec:fss} presents our 
finite-size scaling analysis, split into two parts: 
Section~\ref{sec:static_exponents} covers the extraction 
of the critical temperatures and static critical 
exponents, and Section~\ref{sec:dynamic_exponents} 
covers the estimation of the dynamic critical 
exponents and their comparison between the kagome 
and square lattices. Section~\ref{sec:discussion} 
discusses the physical interpretation of our results. Finally, 
Section~\ref{sec:summary} summarizes our main findings 
and their implications for universality aspects of 
dynamic criticality under cluster algorithms.

\section{Model and lattice geometry}
\label{sec:model}

The site-diluted Ising model is defined via the following Hamiltonian
\begin{equation}
    \mathcal{H} = -J\sum_{\left<ij\right>} \epsilon_i \epsilon_j  \sigma_i \sigma_j\label{eq:hamiltonian},
\end{equation}
where the spin variables take the values $\sigma_i = \pm 1$, 
$J > 0$ is the ferromagnetic exchange coupling constant, 
and the summation runs over all nearest-neighbor pairs 
of spins. The parameter $\epsilon_i$ is $1$ $(0)$ if 
a spin is present (absent) at site $i$, and $p$ denotes 
the probability that a site is occupied, i.e. the spin 
concentration.

\begin{figure}[t]
    \includegraphics[width=0.75\linewidth]{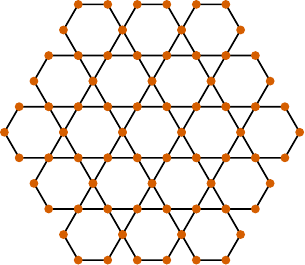}
\caption{Schematic representation of the kagome lattice. 
The lattice consists of corner-sharing triangles alternating 
with hexagonal plaquettes, with each unit cell containing 
three sites forming an equilateral triangle. Each site 
has coordination number four.}
\label{fig:lattice_kagome}
\end{figure}

The kagome lattice is a two-dimensional trihexagonal 
lattice of corner-sharing (but not edge-sharing) 
triangles, consisting exclusively of equivalent sites 
and bonds; see Fig.~\ref{fig:lattice_kagome}. The 
tiling alternates between triangular and hexagonal 
plaquettes, with each site having a coordination 
number of four. The name itself is derived from a 
Japanese woven bamboo basket pattern~\cite{syozi51}. The Ising 
model's transition temperature on this lattice is 
derived from a star-triangle 
transformation~\cite{baxter_book} and is exactly 
$e^{4\beta_{\rm c} J} = 3 + 2\sqrt{3}$~\cite{syozi51}, 
giving $T_{\rm c} \approx 2.14331944\cdots$ in units 
where $k_B = J = 1$, which we adopt throughout this 
work. The exact free energy solution for the Ising 
model on the kagome lattice was obtained by Kan\^{o} 
and Naya~\cite{kano53}, and its spontaneous 
magnetization was subsequently calculated by 
Naya~\cite{naya54}. The site percolation threshold 
of the kagome lattice can be obtained from an 
isomorphism with the bond percolation of the 
honeycomb lattice, giving $p_{\rm c} = 1 - 2\sin{(\pi/18)} 
\approx 0.6527$~\cite{sykes64}, while the bond 
percolation threshold can be found in 
Ref.~\cite{ziff97}. Interest in the kagome lattice 
has remained consistent over the years, with studies 
covering Potts models~\cite{baek11}, 
dipoles~\cite{maksymenko15}, 
dimers~\cite{zeng95, budnik04}, vertex 
models~\cite{zhang26}, tricriticality in the XY 
model~\cite{shahbazi08}, fracton 
excitations~\cite{hering21}, BKT 
transitions~\cite{kakizawa24}, and stacked kagome 
layers~\cite{hemmati12}. The kagome lattice has 
also attracted considerable attention in the context 
of frustrated antiferromagnetic 
interactions~\cite{mila98, waldtmann98, syromyatnikov02, 
nikolic03, jiang08, sindzingre09, soldatov19, zhu25}, 
spin-liquid states~\cite{marston91, hermele05, ran07}, 
and resonating valence bond 
physics~\cite{anderson73, hastings00, yang08, 
evenbly10, singh10}.

One of the main motivations for studying the kagome lattice 
in the context of the diluted Ising model is that it 
provides a geometrically distinct testing ground for 
universality, complementing the more extensively 
studied square lattice. Both lattices are 
two-dimensional and have coordination number four, 
yet their local topology differs fundamentally: 
the square lattice is bipartite with a simple 
plaquette structure, whereas the kagome lattice 
is non-bipartite and consists of corner-sharing 
triangles with three sites per unit cell. This 
geometric difference has no consequence for the 
static universality class, which is determined 
solely by symmetry and dimensionality, but may 
play an important role in the dynamic critical 
behavior under collective update schemes such 
as the Wolff algorithm, where the cluster growth 
is sensitive to the local connectivity of the 
lattice.  A key question we address in 
this work is therefore whether the dynamic critical 
exponent $z$ is shared between the two lattices 
in the pure and diluted cases, and if not, whether 
any observed difference reflects a true asymptotic 
distinction or a finite-size crossover effect driven 
by the different percolation thresholds of the two 
lattices.

\section{Simulations and Observables}
\label{sec:numerics}

We study the site-diluted two-dimensional Ising model on 
the kagome lattice using the Wolff single-cluster 
algorithm~\cite{wolff89,landau_book,newman_book}, 
a collective update scheme that dramatically reduces 
critical slowing down relative to local 
dynamics. The key idea behind this algorithm is to 
build a cluster of spins following the 
Fortuin-Kasteleyn representation~\cite{fortuin72a, 
fortuin72b, fortuin72c} and flip its orientation 
collectively. Specifically, the algorithm proceeds 
as follows:\\
(1) A random spin is chosen as the seed for the cluster and added to a list.\\
(2) Moving through the list, each aligned  neighbor of the current spin that is not already in the cluster is added with probability 
$P_{\rm add} = 1 - e^{-2J\epsilon_i\epsilon_j/T}$.\\
(3) Step 2 is repeated until no further spins can be added to the cluster.\\
(4) All spins in the cluster are flipped 
    simultaneously.\\
\noindent Each such cluster formation and flip 
constitutes one iteration of the algorithm. Since 
the cluster size varies between iterations, the 
Wolff algorithm does not have a standardized time 
step. We therefore define one Monte Carlo sweep 
(MCS) as the average number of cluster-flip 
iterations required to flip $N$ spins in total, 
standardizing the time by a factor 
$f = \langle S \rangle / N$, where $\langle S \rangle$ 
is the average cluster size. We note that the bond acceptance probability $P_{\rm add}$ depends on the critical temperature 
$T_{\rm c}$, which decreases monotonically with 
increasing dilution (see Table~\ref{tab:Tc}), 
leading to a significant increase of $P_{\rm add}$ 
with increasing dilution.

As described in Section~\ref{sec:intro}, see also Fig.~\ref{fig:lattice_kagome}, the kagome 
lattice unit cell consists of three sites forming 
an equilateral triangle.  The total number of lattice sites is therefore $N = 3L^2$, where $L$ is the linear size of the 
underlying Bravais lattice, counting the unit cells 
in each primitive lattice direction. Our large-scale 
simulations include eight linear system sizes, 
$L = \{30, 60, 90, 120, 150, 180, 240, 300\}$, with 
helical boundary conditions imposed throughout. 
We have considered five values of the spin 
concentration: $p = 1$ (pure case) and 
$p = \{0.95, 0.90, 0.85, 0.80\}$, performing simulations 
over $1000$ independent disorder realizations 
for each value of $p$. For each realization, spin 
configurations are initialized in a fully ordered 
state, and $50\,000$ MCS are discarded for thermal 
equilibration before measurements are taken over 
a further $100\,000$ MCS.

A key observable measured during the simulations 
is the time-displaced autocorrelation function of 
an observable $X$, defined at time $t$ as
\begin{equation}
    \label{eq:autocorrelation_function}
    C_X(t) = \left\langle \left(X_i - \langle X \rangle\right) 
    \left(X_{i+t} - \langle X \rangle\right) \right\rangle,
\end{equation}
where $\langle \cdots \rangle$ denotes thermal averaging.
For practical reasons, the autocorrelation functions 
are computed using the number of cluster moves as 
the unit of time, which is subsequently rescaled 
to MCS by the factor $f$, 
as described above. For large $t$, the normalized autocorrelation function 
decays exponentially~\cite{janke12},
\begin{equation}
    \label{eq:tau_exp}
    \frac{C_X(t)}{C_X(0)} \propto e^{-t/\tau_{X,\mathrm{exp}}},
\end{equation}
from which the exponential autocorrelation time 
$\tau_{X,\mathrm{exp}}$ can be extracted. The 
integrated autocorrelation time is defined as
\begin{equation}
    \label{eq:tau_int}
    \tau_{X,\mathrm{int}} = \frac{1}{2} + 
    \sum_{t=1}^{\infty} \frac{C_X(t)}{C_X(0)}.
\end{equation}
Due to the finite length of our time series, the 
direct numerical evaluation of Eq.~\eqref{eq:tau_int} 
suffers from systematic errors at large $t$, arising 
from the accumulation of noise in the tail of the 
autocorrelation function. To address this, we 
follow the standard approach of Ref.~\cite{janke12}: 
the sum is truncated at a time $t_{\rm max}$ once 
the self-consistent cut-off condition 
$t_{\rm max} \geq 6\,\tau_{X,\mathrm{int}}$ is 
satisfied~\cite{madras88}, and the contribution 
from the truncated tail is recovered analytically 
using the exponential autocorrelation 
time~\cite{janke95},
\begin{equation}
    \label{eq:tau_int_with_trailing_tau_exp}
    \tau_{X,\mathrm{int}}(t_{\rm max}) = 
    \tau_{X,\mathrm{int}} - a\, 
    \frac{e^{-t_{\rm max}/\tau_{X,\mathrm{exp}}}}
    {e^{1/\tau_{X,\mathrm{exp}}} - 1},
\end{equation}
where $a$ is the proportionality constant of 
Eq.~\eqref{eq:tau_exp}. For the disordered cases 
($p < 1$), we compute the quenched disorder average 
of the integrated autocorrelation time following 
the procedure of Ref.~\cite{janke00}, defining 
the final disorder-averaged autocorrelation time 
as $\tau_{X,\mathrm{T}} = [f] \cdot [\tau_{X,\mathrm{T}}]$, 
where the subscript $\mathrm{T}$ denotes either 
$\mathrm{int}$ or $\mathrm{exp}$, and square 
brackets denote averaging over disorder realizations.
,
\begin{figure}[t]
\centering
\includegraphics[width=\linewidth]{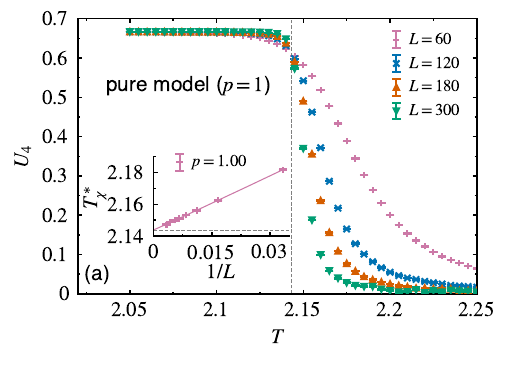}
\centering
\includegraphics[width=\linewidth]{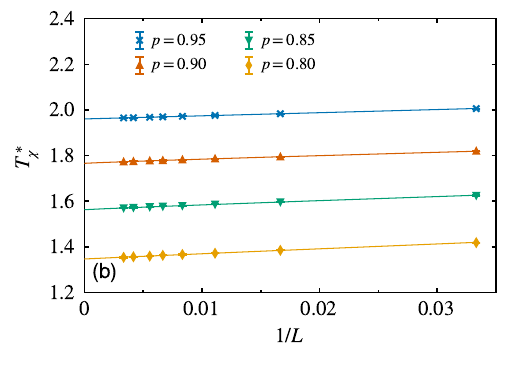}
\caption{(a) Fourth-order Binder cumulant $U_4$ versus temperature for the pure kagome Ising model ($p = 1$) and characteristic system sizes. The crossing of the curves locates the critical temperature $T_{\rm c}$ (vertical dashed line). The inset shows the finite-size scaling of the $\chi$ maxima locations $T^{\ast}_\chi$. (b) Equivalent scaling of the pseudo-critical temperatures $T^{\ast}_{\chi}$ for the site-diluted model, for all values of $p$ considered.}
    \label{fig:T_scaling}
\end{figure}

\begin{figure}[t]
\centering
\centering
\includegraphics[width=\linewidth]{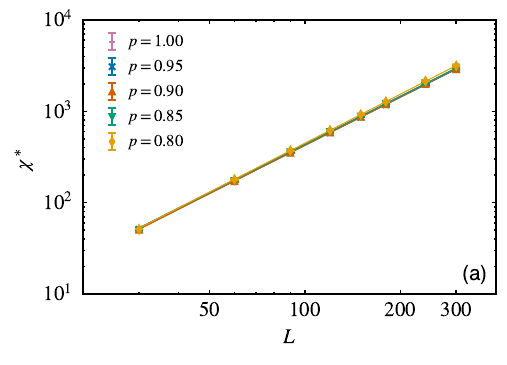}
\centering
\includegraphics[width=\linewidth]{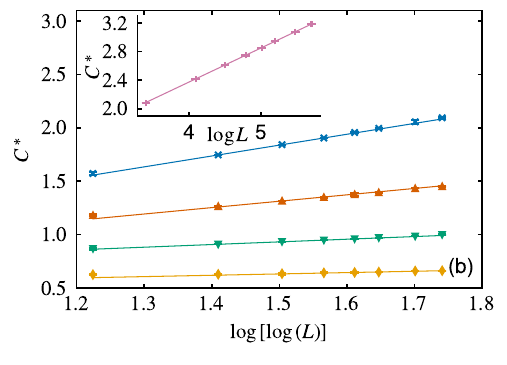}
\caption{(a) Finite-size scaling of the magnetic susceptibility peaks $\chi^{\ast}$ for the site-diluted Ising model on the kagome lattice, for all values of the spin concentration $p$ considered. The solid lines represent fits to the power-law ansatz $\chi^{\ast} \sim L^{\gamma/\nu}$, with the extracted exponent ratios $\gamma/\nu$ consistent with the exact Ising value $7/4$ for all $p$ (see Table~\ref{tab:gamma_nu}). (b) 
Finite-size scaling of the specific heat peak $C^{\ast}$ 
on the kagome lattice. The main panel shows the 
double-logarithmic scaling $C^{\ast} \sim \log[\log(L)]$ 
observed for the site-diluted cases ($p < 1$), while 
the inset demonstrates the logarithmic scaling 
$C^{\ast} \sim \log(L)$ expected for the pure ($p = 1$) Ising 
universality class. }
\label{fig:C_chi}
\end{figure}

For every $\tau_{X,\mathrm{T}}$ there is a corresponding dynamical critical exponent $z_{X,\mathrm{T}}$, which governs the critical slowing down via the power-law relation $\tau_{X,\mathrm{T}} \sim L^{z_{X,\mathrm{T}}}$ at criticality~\cite{newman_book}. In order to study the subtle dependence of $z_{X,\mathrm{T}}$ 
on the dilution strength $p$, we resort to a more refined 
approach based on effective exponents estimated from pairs 
of system sizes $(L, L^\prime)$. Specifically, from the 
power-law scaling of $\tau_{X,\mathrm{T}} \sim L^{z_{X,\mathrm{T}}}$, 
an effective exponent can be defined as
\begin{equation}
z_{X,\mathrm{T}}^{(\mathrm{eff})}(L, L^\prime) = 
\frac{\log\left[\tau_{X,\mathrm{T}}(L^\prime) / 
\tau_{X,\mathrm{T}}(L)\right]}{\log\left[L^\prime / L\right]},
\end{equation}
whose value approaches the true asymptotic exponent 
$z_{X,\mathrm{T}}$ in the limit $L, L^\prime \to \infty$, 
up to corrections to scaling that vanish with increasing 
system size. Extrapolating $z_{X,\mathrm{T}}^{(\mathrm{eff})}$ 
to the thermodynamic limit therefore 
provides a more controlled and precise estimate of 
$z_{X,\mathrm{T}}$ than a direct power-law fit, and allows 
the subtle variation of $z$ with $p$ to be resolved more 
clearly. For consistency and to reduce the number of 
independent pairs, we constrain the analysis throughout 
to $L^\prime = 2L$.

\begin{table}[t]
\caption{Critical temperatures $T_{\rm c}$, thermal 
critical exponent $\nu$, minimum system size 
$L_{\rm min}$, and goodness-of-fit parameter $Q$ 
for the site-diluted kagome Ising model, for all 
values of $p$ considered. The values of $\nu$ are 
consistent with the exact Ising value $\nu = 1$ 
for all dilutions. The Wolff bond acceptance 
probability $P_{\rm add} = 1 - e^{-2J/T_{\rm c}}$ 
at criticality increases monotonically from 
$0.607$ at $p = 1$ to $0.774$ at $p = 0.80$, 
reflecting the reduction of $T_{\rm c}$ with 
increasing dilution. See also Fig.~\ref{fig:T_scaling}.}
\label{tab:Tc}
\vspace*{0.1cm}
\setlength{\tabcolsep}{4pt}
\centering
\begin{tabular}{ccccc}
\hline\hline
$p$ & $T_{\rm c}$ & $\nu$ & $L_{\rm min}$ & $Q$ \\[0.5ex]
\hline\hline
$1.00$ & $2.1439(5)$ & $0.98(3)$ & $30$ & $51\%$ \\[0.5ex]
\hline
$0.95$ & $1.9605(3)$ & $0.99(3)$ & $60$ & $11\%$ \\[0.5ex]
\hline
$0.90$ & $1.7659(6)$ & $1.06(12)$ & $60$ & $21\%$ \\[0.5ex]
\hline
$0.85$ & $1.5626(13)$ & $1.09(9)$ & $90$ & $40\%$ \\[0.5ex]
\hline
$0.80$ & $1.3465(21)$ & $1.05(10)$ & $60$ & $15\%$ \\[0.5ex]
\hline\hline
\end{tabular}
\end{table}

The primary observables for which we compute 
autocorrelation times are the internal energy 
per site $E = \langle H \rangle / N$ and the 
magnetization per site $M = \frac{1}{N}\sum_{i=1}^N 
\epsilon_i \sigma_i$. From these we also compute 
the specific heat $C = \beta^2 N\left[\langle E^2 
\rangle - \langle E \rangle^2\right]$ and the 
magnetic susceptibility $\chi = \beta N\left[\langle 
M^2 \rangle - \langle M \rangle^2\right]$, which 
are used in the finite-size scaling analysis of 
Section~\ref{sec:static_exponents}. As a consistency 
check, we also measure the fourth-order Binder 
cumulant of the magnetization, $U_4 = 1 - \langle M^4 \rangle /[3\langle M^2 \rangle^2]$,
whose crossing point as a function of temperature 
for different system sizes provides an independent 
estimate of the critical temperature.

Statistical errors are estimated using the jackknife 
procedure after binning the data into $100$ groups. 
Critical exponents are then determined by weighted 
least-squares fits to the data from different system 
sizes. A fit is considered acceptable if the 
goodness-of-fit parameter satisfies $10\% \leq Q 
\leq 90\%$ and the reduced chi-squared is close to 
unity. To minimize finite-size effects, fits are 
performed using only system sizes $L \geq L_{\rm min}$, 
where $L_{\rm min}$ is chosen separately for each 
observable and dilution value, and is reported 
alongside the fit results in the tables of 
Section~\ref{sec:fss}.

\section{Finite-size scaling analysis}
\label{sec:fss}

\subsection{Static Ising universality}
\label{sec:static_exponents}

We begin our finite-size scaling analysis with the 
extraction of the critical temperatures $T_{\rm c}(p)$ 
for all values of the spin concentration $p$ considered. 
Following the standard scaling ansatz~\cite{newman_book}
\begin{equation}
\label{eq:Tc_scaling}
T^{\ast}_{\chi} = T_{\rm c} + \mathcal{A} L^{-1/\nu},
\end{equation}
where $T^{\ast}_{\chi}$ denotes the location of the $\chi$-curve peak for system size $L$ and dilution $p$, 
$\mathcal{A}$ corresponds to non-universal scaling amplitudes, and $\nu$ is the thermal critical exponent associated 
with the correlation length, we fit the pseudo-critical 
temperatures extracted from the peaks of the magnetic 
susceptibility $\chi$. The results of this analysis 
are shown in Fig.~\ref{fig:T_scaling} and summarized 
in Table~\ref{tab:Tc}. 
Figure~\ref{fig:T_scaling}(a) shows the fourth-order 
Binder cumulant $U_4$ for the pure case ($p = 1$), 
whose crossing in close proximity to the critical 
point (vertical dashed line) provides an independent 
confirmation of $T_{\rm c}$. The inset illustrates 
the finite-size scaling of the $\chi$ maxima for 
$p = 1$, while Fig.~\ref{fig:T_scaling}(b) shows the equivalent 
scaling for all diluted cases ($p < 1$). For $p = 1$, 
our estimate of the critical temperature is in 
perfect agreement with the exact value~\cite{syozi51}, $T_{\rm c} \approx 2.14331944\cdots$. For the diluted cases, $T_{\rm c}$ decreases monotonically with increasing dilution, leading 
to a significant increase of the bond acceptance 
probability $P_{\rm add} = 1 - e^{-2J/T_{\rm c}}$ 
in the Wolff cluster. The significant increase of $P_{\rm add}$ with decreasing $p$---from $\sim 0.607$ at $p = 1$ to 
$\sim 0.774$ at $p = 0.80$---has important consequences 
for the Wolff cluster dynamics, as we discuss in 
Sec.~\ref{sec:dynamic_exponents}. The values of the 
thermal critical exponent $\nu$ extracted from the 
fits of Eq.~\eqref{eq:Tc_scaling} are reported in 
Tab.~\ref{tab:Tc} for all values of $p$. In all 
cases, the results are consistent with the exact 
pure Ising value $\nu = 1$, in agreement with the strong universality hypothesis~\cite{dotsenko83,shankar87,ludwig87, ludwig87b,ludwig90,wang90,shalaev94}. 

\begin{table}[t]
\caption{Estimates of $\gamma/\nu$ from fits of the 
magnetic susceptibility peaks $\chi^{\ast} \sim L^{\gamma/\nu}$ 
for the site-diluted kagome Ising model, for all values 
of $p$ considered, together with the corresponding 
$L_{\rm min}$ and goodness-of-fit parameter $Q$. Fit 
qualities are also reported for the logarithmic 
($p = 1$) and double-logarithmic ($p < 1$) scaling 
of the specific heat peak. All estimates of $\gamma/\nu$ 
are consistent with the exact Ising value $7/4$; 
see also Fig.~\ref{fig:C_chi}.}
\label{tab:gamma_nu}
\vspace*{0.2cm}
\centering
\begin{tabular}{cccccc}
\hline \hline
\multicolumn{1}{c}{}    & \multicolumn{3}{c}{magnetic susceptibility ($\chi$)}                                                  & \multicolumn{2}{l}{specific heat ($C$)} \\
\multicolumn{1}{c}{$p$} & \multicolumn{1}{c}{$\gamma/\nu$} & \multicolumn{1}{c}{$L_{\rm min}$} & \multicolumn{1}{c}{$Q$} & $L_{\rm min}$       & $Q$        \\ \hline
$1.00$                  & $1.75(1)$                        & $30$                              & $45\%$                & $60$                & $11\%$      \\ [0.5ex]
\hline
$0.95$                  & $1.75(1)$                        & $60$                              & $12\%$                & $60$                & $20\%$      \\[0.5ex]
\hline
$0.90$                  & $1.76(2)$                        & $120$                             & $50\%$                & $90$                & $23\%$      \\[0.5ex]
\hline
$0.85$                  & $1.75(1)$                        & $120$                             & $51\%$                & $60$                & $11\%$      \\[0.5ex]
\hline
$0.80$                  & $1.77(2)$                        & $60$                              & $32\%$                & $60$                & $12\%$ \\ [0.5ex]
\hline \hline
\end{tabular}
\end{table}

We next examine the scaling of the magnetic  susceptibility and specific heat peaks, shown in Fig.~\ref{fig:C_chi} and summarized in Table~\ref{tab:gamma_nu}. Figure~\ref{fig:C_chi}(a) demonstrates that the peak values of $\chi$ follow the well-known power-law scaling with system size $\chi^{\ast} \sim L^{\gamma/\nu}$,
irrespective of the dilution strength. Fits to  this ansatz yield exponent ratios $\gamma/\nu$  in excellent agreement with the exact Ising  universality class value of $7/4$ for all $p$ considered (see Table~\ref{tab:gamma_nu}), providing further confirmation that the static universality class is preserved under site dilution. Figure~\ref{fig:C_chi}(b) shows the scaling of the specific heat peak $C^{\ast}$ with system size. For the pure case ($p = 1$, inset), the specific heat peak grows logarithmically 
with system size, $C^{\ast} \sim \log(L)$, as expected 
for the Ising universality class. Upon the introduction of any nonzero dilution ($p < 1$, main panel), this growth slows to a 
double-logarithmic dependence, $C^{\ast} \sim 
\log[\log(L)]$, consistent with the multiplicative 
logarithmic corrections to scaling predicted theoretically~\cite{dotsenko83,shankar87,ludwig90}. All fit qualities are  within the accepted range and are reported in Table~\ref{tab:gamma_nu}. 

\begin{figure}[t]
    \centering
    \includegraphics[width=\linewidth]{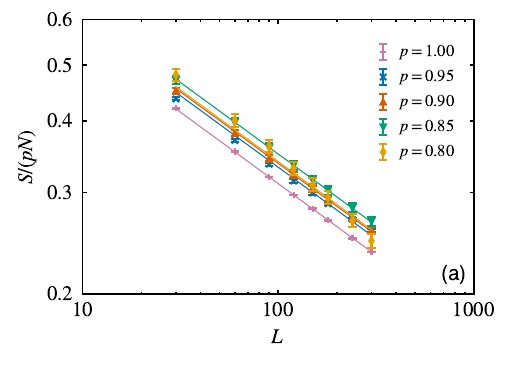}
    \includegraphics[width=\linewidth]{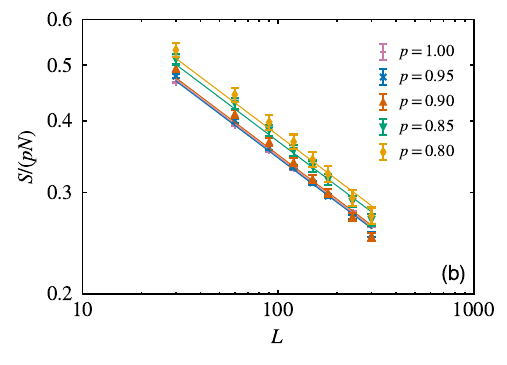}
    \caption{Scaling of the normalized average Wolff cluster size $S/(pN)$ with system size $L$ for the site-diluted  Ising model on the kagome (a) and square (b) lattices, for all values of the spin  concentration $p$ considered. Solid lines are power-law fits of the form $S \sim L^{\gamma/\nu}$, yielding exponent ratios that average to $\gamma/\nu = 1.747(4)$ across all dilutions. Both axes are on a logarithmic scale, with the 
power-law scaling appearing as straight lines.}
\label{fig:cluster_size_square} 
\end{figure}

As an additional and independent test of the  static universality, we present in Fig.~\ref{fig:cluster_size_square} the scaling of the normalized average Wolff cluster size $S/(pN)$ with system size $L$, for both the kagome and square Ising models across all values of $p$. Since  the average cluster size is proportional to  the magnetic susceptibility, $S \propto \chi$,  it is expected to scale as $S \sim L^{\gamma/\nu}$  at criticality~\cite{stauffer_book}. Fits to this power law yield $\gamma/\nu = 1.747(4)$  across all dilutions and on both lattices, in  excellent agreement with the exact Ising value  of $7/4$. Taken together, the results of this section  paint a consistent and complete picture: the  static universality class of the two-dimensional Ising model is fully preserved under site dilution on the kagome lattice, with critical  exponents indistinguishable from their exact  pure values, and is shared between the kagome and square lattices across the full range of dilutions considered. This sets the stage for  the dynamic analysis of the following section, where a different picture emerges.

\subsection{Dynamic critical exponents}
\label{sec:dynamic_exponents}

\begin{figure}[t]
    \centering
    \includegraphics[width=\linewidth]{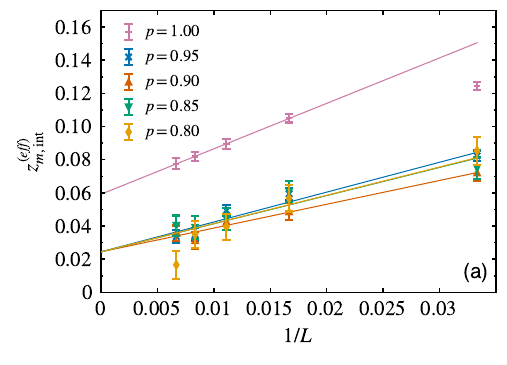}
    \includegraphics[width=\linewidth]{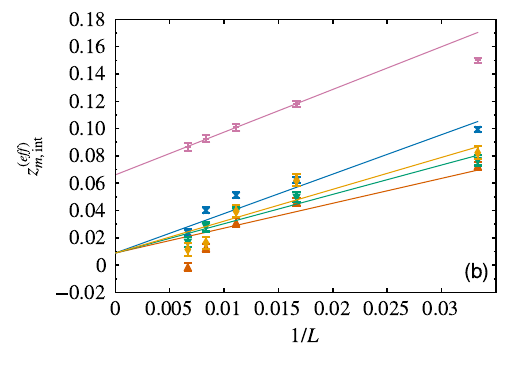}
\caption{Effective dynamic critical exponent 
$z_{m, \rm{int}}^{(\mathrm{eff})}$ versus $1/L$ for the 
site-diluted kagome (a) and square 
(b) Ising models, for all values 
of $p$ considered. Solid lines are fits to the 
thermodynamic limit.}
    \label{fig:zeff_m}
\end{figure}

\begin{figure}[t]
    \centering
    \includegraphics[width=\linewidth]{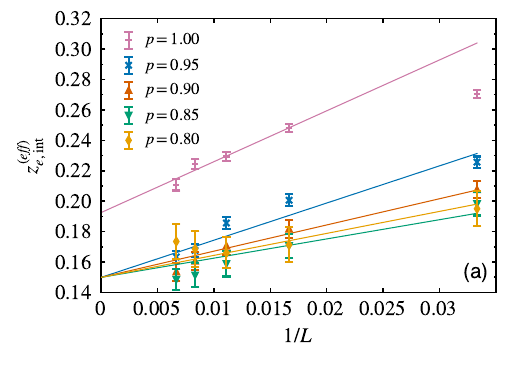}
    \includegraphics[width=\linewidth]{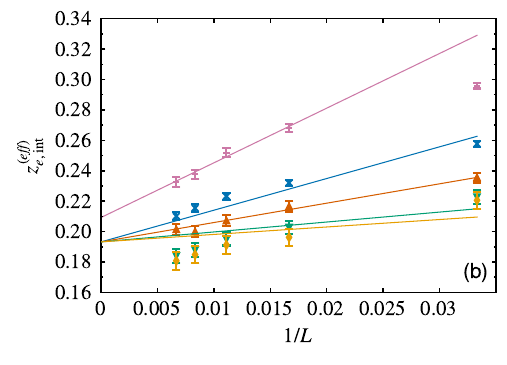}
\caption{Same as Fig.~\ref{fig:zeff_m}, but for 
the effective dynamic critical exponent of the 
energy, $z_{e}^{(\mathrm{eff})}$. The energy-based exponent is systematically larger than the magnetization-based exponent shown in Fig.~\ref{fig:zeff_m}, consistent with the Wolff algorithm more effectively reducing critical slowing down for the order parameter.}
    \label{fig:zeff_e}
\end{figure}

We now turn to the dynamic analysis, based on the 
effective exponent approach described in 
Section~\ref{sec:numerics}. Figures~\ref{fig:zeff_m} 
and~\ref{fig:zeff_e} show the effective dynamic 
critical exponents $z_{m,\mathrm{int}}^{(\mathrm{eff})}$ 
and $z_{e,\mathrm{int}}^{(\mathrm{eff})}$, extracted 
from the magnetization and energy autocorrelation 
times respectively, as a function of $1/L$ for both 
the kagome and square lattices and all values of $p$ 
considered. In both cases, the thermodynamic limit 
extrapolation reveals a clear separation between 
the pure case ($p = 1$) and the diluted cases 
($p < 1$), which we discuss in turn; see also Table~\ref{tab:dynamic_exponents}.

For the magnetization-based exponent, shown in 
Fig.~\ref{fig:zeff_m}, the pure case yields 
$z_{m,\mathrm{int}} = 0.059(5)$ for the kagome 
lattice and $z_{m,\mathrm{int}} = 0.066(4)$ for 
the square lattice, giving a combined estimate of 
$z_{m,\mathrm{int}} = 0.063(3)$. These values are 
consistent within statistical errors, confirming 
that dynamic universality is shared between the 
two lattices in the pure case. For $p < 1$, joint 
fits over all diluted cases yield 
$z_{m,\mathrm{int}} = 0.024(2)$ for the kagome 
lattice and $z_{m,\mathrm{int}} = 0.009(2)$ for 
the square lattice. Both values are significantly 
smaller than their respective pure-case estimates, 
in agreement with previous studies of site- and 
bond-diluted systems under cluster 
dynamics~\cite{hennecke93,ivaneyko06,kanbur24,vatansever25}, 
where an increase in randomness strength was found 
to decrease $z$. However, the two 
lattices yield significantly different values for 
$p < 1$, in contrast to the pure case, 
indicating that the introduction of site dilution leads to lattice-dependent effective values of $z$, consistent with the crossover picture discussed below.

The same picture emerges from the energy-based 
exponent, shown in Fig.~\ref{fig:zeff_e}. For 
$p = 1$, the kagome and square lattices yield 
$z_{e,\mathrm{int}} = 0.198(6)$ and 
$z_{e,\mathrm{int}} = 0.209(6)$ respectively, 
giving a combined estimate of 
$z_{e,\mathrm{int}} = 0.204(4)$, consistent within 
errors and again confirming dynamic universality 
in the pure case. For $p < 1$, joint fits yield 
$z_{e,\mathrm{int}} = 0.150(2)$ for the kagome 
lattice and $z_{e,\mathrm{int}} = 0.193(2)$ for 
the square lattice, values that are both 
reduced relative to the pure case 
and significantly different from each other,
reinforcing the observation that the two lattices 
yield different effective values of $z$ in the diluted 
regime, consistent with a dilution-induced crossover 
between the Ising and percolation fixed points. We note that across all cases, 
$z_{e,\mathrm{int}}$ is significantly larger 
than $z_{m,\mathrm{int}}$, consistent with the 
Wolff algorithm more effectively reducing 
critical slowing down for the magnetization 
than for the energy. We therefore regard 
$z_{e,\mathrm{int}}$ as the more robust 
indicator of dynamic universality, and base 
our main conclusions on this quantity. 

Closing this section, we comment briefly on corrections to scaling. All scaling ansätze used throughout this work retain 
only the leading scaling behavior, neglecting subleading 
corrections of the form $(1 + bL^{-\omega})$. For the 
pure two-dimensional Ising model, $\omega = 1.75$~\cite{blote:88,shao:16}, implying that such corrections 
decay rapidly and are negligible at our system sizes. 
For the dynamic scaling in the diluted cases, we 
additionally explored corrections of the form 
$\sim \log(L)/L$, motivated by the double-logarithmic 
scaling of the specific heat. In all cases---static 
and dynamic, pure and diluted---the explicit inclusion 
of correction-to-scaling terms yielded no statistically 
significant improvement in fit quality, and all 
extracted exponents remained consistent with the 
leading-order results within statistical errors. 
The systematic use of a minimum system size $L_{\rm min}$ 
in all fits provides an additional safeguard against 
residual finite-size effects.

\section{Discussion}
\label{sec:discussion}

The numerical results of Section~\ref{sec:fss} 
establish a clear contrast between the static and 
dynamic critical behavior of the site-diluted kagome 
Ising model under Wolff dynamics. In this section 
we discuss the physical interpretation of these 
results, addressing in turn the mechanism behind 
the reduction of $z$ with dilution, the origin of 
the lattice-dependent crossover behavior observed 
in the diluted regime, and the role of rare-region 
effects and corrections to scaling.

It is instructive at this point to provide a general physical argument for the observed reduction of the dynamic critical 
exponent $z$ under Wolff dynamics upon the introduction 
of site dilution. In the pure system, the Wolff algorithm 
builds Fortuin-Kasteleyn clusters whose average size scales as 
$\langle S \rangle \sim L^{\gamma/\nu}$ at criticality, 
and the autocorrelation time measures how many such 
cluster flips are required to decorrelate the full 
system of $N$ spins. Upon the introduction of site 
dilution, two competing effects modify this picture. 
On the one hand, the increase of $P_{\rm add}$ with 
decreasing $p$ tends to produce larger clusters, 
which would naively increase the decorrelation 
efficiency. On the other hand, the fragmentation of 
the lattice connectivity by vacancies prevents the 
clusters from spanning the system as effectively 
as in the pure case, bounding their growth by the 
local connectivity structure. The net result is 
that the Wolff clusters in the diluted system are 
more targeted: rather than flipping large, system-spanning 
clusters of the full lattice, the algorithm flips 
clusters that are large relative to the local 
connected regions of the active sublattice. Since 
the autocorrelation time is normalized by the 
fraction of active spins through the factor 
$f$, and since the 
effective number of spins that need to be 
decorrelated is reduced from $N$ to $pN$, each 
cluster flip decorrelates a larger fraction of 
the magnetically active system per unit of 
computational effort. 
This increased efficiency 
of the Wolff algorithm relative to the active 
sublattice is the geometric origin of the 
reduction of $z$ with dilution: the algorithm 
effectively solves a smaller problem with each 
flip, and the critical slowing down, measured 
in units of MCS, is correspondingly reduced. 
This argument also explains why the effect is 
in the opposite direction to that observed under 
local dynamics, where each spin 
flip targets a single site regardless of the 
cluster structure, and disorder simply introduces 
additional barriers to equilibration that increase 
$z$. The cluster algorithm, by contrast, 
automatically adapts its update to the local 
connectivity of the diluted lattice, turning 
the disorder from an obstacle into an advantage 
for efficient sampling.

A central result of our dynamic analysis is that while 
$z$ is consistent between the kagome and square lattices 
in the pure case, it differs between the two lattices 
upon the introduction of site dilution. To understand 
this, we note that the renormalization-group flow of 
the diluted Ising model is governed by two fixed points: 
the pure Ising fixed point at $p = 1$ and the 
percolation fixed point at $p = p_{\rm c}$. A compelling 
geometric argument connecting these two fixed points 
is provided by the fractal structure of the clusters 
flipped by the Wolff algorithm. In the pure Ising model 
at criticality, the Wolff algorithm flips Fortuin-Kasteleyn clusters with fractal dimension $d_f = 2-\beta/\nu = 15/8$, which 
determines how efficiently each cluster flip decorrelates 
the system and hence governs $z$. As the dilution 
strength increases and $p \to p_{\rm c}^+$, the clusters 
relevant to the Wolff dynamics are no longer pure Ising 
Fortuin-Kasteleyn clusters but increasingly resemble percolation 
clusters, whose fractal dimension in two dimensions is 
$d_f^{\rm (perc)} = 91/48$~\cite{stauffer_book}. Since 
$91/48 \neq 15/8$, the geometric structure of the 
clusters flipped by the algorithm changes continuously 
as $p$ decreases from $1$ toward $p_{\rm c}$, becoming 
increasingly ramified and percolation-like, and $z$ 
smoothly crosses over from its pure Ising value toward 
the value characteristic of percolation-dominated 
dynamics. 

\begin{table}[t]
\caption{Estimates of the dynamic critical exponent 
extracted from the integrated autocorrelation times 
of the magnetization and energy under Wolff 
single-cluster dynamics, for the site-diluted Ising 
model on the kagome and square lattices, for all 
values of the spin concentration $p$ considered. For $p = 1$ 
the values are obtained from linear extrapolations 
of $z^{(\mathrm{eff})}$ as a function of $1/L$, 
while for $p_{\rm c} < p < 1$ the values are obtained from 
joint fits over all diluted cases; see Figs.~\ref{fig:zeff_m} and~\ref{fig:zeff_e}. The combined 
estimates for the pure case are $z_{m,\mathrm{int}} = 0.063(3)$ and $z_{e,\mathrm{int}} = 0.204(4)$, obtained by averaging the kagome and square lattice values in quadrature.}
\label{tab:dynamic_exponents}
\vspace*{0.1cm}
\centering
\begin{tabular}{ccc}
\hline\hline
$p$ & $z_{m,\mathrm{int}}$ & $z_{e,\mathrm{int}}$ \\[0.5ex]
\hline\hline
\multicolumn{3}{c}{kagome lattice} \\
\hline
$1$ & $0.059(5)$ & $0.198(6)$ \\[0.5ex]
$p_{\rm c} < p < 1$ & $0.024(2)$ & $0.150(2)$ \\[0.5ex]
\hline\hline
\multicolumn{3}{c}{square lattice} \\
\hline
$1$ & $0.066(4)$ & $0.209(6)$ \\[0.5ex]
$p_{\rm c} < p < 1$ & $0.009(2)$ & $0.193(2)$ \\[0.5ex]
\hline\hline
\end{tabular}
\end{table}

Crucially, since percolation universality 
is determined by dimension alone and not by lattice 
geometry, all percolation exponents---including the 
dynamic exponents and the cluster fractal dimension 
$d_f^{\rm (perc)} = 91/48$---are identical for the 
kagome and square lattices in two dimensions. 
Similarly, as our data confirm, the dynamic exponent 
$z$ at $p = 1$ is also shared between the two lattices. 
The renormalization-group flow therefore connects the 
same two fixed points, with the same dynamic exponents 
and the same cluster fractal dimensions, on both 
lattices. The only difference between the two lattices 
is the value of $p_{\rm c}$: $p_{\rm c} \approx 0.6527$ 
for the kagome lattice~\cite{sykes64} and 
$p_{\rm c} \approx 0.5927$ for the square 
lattice~\cite{stauffer_book}. As a consequence, for 
a given value of $p \in (p_{\rm c}, 1)$, the kagome 
lattice sits effectively closer to its percolation 
threshold than the square lattice, and therefore at 
a different position along the same crossover 
trajectory, yielding a different effective $z$ at 
the system sizes accessible in our simulations. The 
observed difference in the effective values of $z$ 
between the two lattices is therefore most naturally 
interpreted as a finite-size crossover effect, rather 
than a true asymptotic distinction between different 
dynamic universality classes. In this picture, the 
true asymptotic dynamic exponent, accessible only 
at much larger system sizes or at values of $p$ well 
above both percolation thresholds, should ultimately 
be shared between the two lattices, consistent with 
the theoretical expectation that dynamic universality 
is determined by dimension and symmetry rather than 
by microscopic lattice geometry. Confirming this 
picture definitively requires either simulations at 
significantly larger system sizes or a systematic 
study of $z$ as $p \to p_{\rm c}^+$ on both lattices, 
which we leave for future work.

We note also that Kole \emph{et al.}~\cite{kole22} demonstrated 
that in the diluted case, the Wolff algorithm can suffer 
from anomalously long correlation times caused by isolated 
(groups of) spins that are infrequently visited by the 
cluster growth. They showed that such rare regions provide 
a lower bound on the autocorrelation time of the form 
$\tau \gtrsim L^{z_w}$, with a dynamical exponent 
$z_w = \gamma/\nu = 7/4$, for any bond concentration 
$p < 1$. This bound is significantly larger than the 
values of $z_{e,\mathrm{int}}$ and $z_{m,\mathrm{int}}$ 
extracted in our simulations, which might appear 
contradictory. However, the density of such isolated 
clusters is extremely small for the values of $p$ 
considered in this work, all of which are well above 
the percolation threshold, and their contribution to 
the autocorrelation time is therefore negligible at 
the system sizes we simulate. We thus expect this 
rare-region mechanism to become relevant only for 
values of $p$ much closer to $p_{\rm c}$, beyond the range 
explored here.

\section{Summary}
\label{sec:summary}

We have studied the static and dynamic critical behavior 
of the site-diluted Ising model on the kagome lattice 
using the Wolff single-cluster algorithm~\cite{wolff89}, 
for five values of the spin concentration $p$ and system 
sizes up to $L = 300$. On the static side, the thermal 
exponent $\nu = 1$ and the exponent ratio $\gamma/\nu = 7/4$ 
retain their exact pure Ising values for all $p$ considered, 
with the specific heat peak growing logarithmically with 
system size for $p = 1$ and crossing over to a 
double-logarithmic growth for $p < 1$, consistent with 
the multiplicative logarithmic corrections predicted by 
theory~\cite{dotsenko83, shankar87, ludwig87, ludwig87b, ludwig90, wang90, shalaev94}. These results are 
shared between the kagome and square lattices, confirming 
that the static universality class is lattice-independent 
across all dilutions considered. On the dynamic side, the picture is fundamentally different. The dynamic critical exponent $z$ decreases upon the introduction of site dilution 
on both lattices, opposite to what is observed under 
local dynamics~\cite{hasenbusch07}, and 
consistent with previous results for bond-diluted 
systems~\cite{kanbur24,vatansever25}. Additionally, it shows 
evidence of saturating to a $p$-independent value at 
strong dilution, suggesting a diluted dynamic fixed 
point. Most strikingly, while $z$ is consistent between 
the kagome and square lattices in the pure case, 
it differs in the diluted regime, an effect we 
attribute to a finite-size crossover driven by 
the different percolation thresholds of the two 
lattices, rather than a true asymptotic breakdown 
of dynamic universality. We note also that the magnetization-based exponent $z_m$ is 
significantly smaller than $z_e$ across all cases, 
consistent with the Wolff algorithm nearly eliminating 
critical slowing down for the order parameter, and we 
regard $z_e$ as the more robust indicator of dynamic 
universality. Taken together, our results point toward a physically rich picture in which static and dynamic universality respond differently to the interplay between site 
dilution and lattice geometry: while the former is 
fully preserved and lattice-independent for all $p$, 
the latter, though shared between the two lattices 
in the pure case, shows lattice-dependent behavior 
in the diluted regime consistent with a crossover 
between the Ising and percolation fixed points.
Settling this question definitively, and developing a theoretical framework capable of describing cluster dynamics in disordered systems, remain important challenges for future work.

\begin{acknowledgments}
The numerical calculations reported in this paper were performed at T\"{U}B\.{I}TAK ULAKBIM (Turkish agency), High Performance and Grid Computing Center (TRUBA Resources). The work of AV and NGF was supported by the Leverhulme Trust through a Leverhulme Trust Research 
Project Grant (grant RPG-2026-025).
\end{acknowledgments}

\bibliography{biblio}

\end{document}